\documentstyle[12pt]{article}

\def\be{\begin{equation}}
\def\ee{\end{equation}}
\def\bea{\begin{eqnarray}}
\def\eea{\end{eqnarray}}

\def\br{\begin{thebibliography}{abc}}
\def\er{\end{thebibliography}}

\def\a{\alpha}
\def\b{\beta}

\def\d{\delta}
\def\e{\epsilon}
\def\f{\phi}
\def\g{\gamma}
\def\h{\eta}

\def\j{\psi}

\def\l{\lambda}
\def\m{\mu}
\def\n{\nu}
\def\o{\omega}
\def\p{\pi}
\def\q{\theta}
\def\r{\rho}

\def\F{\Phi}

\def\L{\Lambda}
\def\O{\Omega}
\def\P{\Pi}

\def\cd{{\cal D}}

\def\cg{{\cal G}}
\def\ch{{\cal H}}

\def\cm{{\cal M}}
\def\cn{{\cal N}}

\def\pa{\partial}

\def\bar#1{\overline{#1}}

\def\Hat#1{\rlap{\kern.10em$\widehat{\phantom G}$}#1}
\def\HAt#1{\rlap{\kern.05em$\widehat{\phantom G}$}#1}

\def\czp#1{\rlap{\kern.1em$\widehat{\phantom{G\vrule height.8em}}$}#1{}}
\def\Czp#1{\rlap{\kern.05em$\widehat{\phantom{G\vrule height.8em}}$}#1{}}

\def\amd{A_{\mu}}
\def\amu{A^{\mu}}
\def\aid{A_{i}}

\def\fmd{F_{\mu \nu}}
\def\fmu{F^{\mu \nu}}

\def\ma{m_{A}}
\def\pnm{\Psi_{nm}}
\def\ponm{\Psi_{nm}^{(0)}}
\def\hn{h_{n}}
\def\wn{\o_{nm}}
\def\qn{q_{nm}}
\def\qon{q_{nm}^{(0)}}

\def\pn{p_{nm}}
\def\pon{p_{nm}^{(0)}}

\def\fnm{\Phi_{nm}}
\def\fonm{\Phi_{nm}^{(0)}}
\def\Hn{H_{n}}
\newcommand{\sect}[1]{\setcounter{equation}{0}\section{#1}}

\def\sxn#1{\bigskip\medskip \sect{#1} \smallskip
                                                 }

\begin{document}

\thispagestyle{empty}
\setcounter{page}{0}

\begin{flushright}
SU-4240-575\\
August 1994
\end{flushright}
\vspace*{15mm}
\centerline {\LARGE Inequivalence of the Massive Vector Meson and}
\vspace{5mm}
\centerline {\LARGE Higgs Models on a Manifold with Boundary}
\vspace*{15mm}
\centerline {\large L. Chandar and E. Ercolessi\footnote{Present address:
Dipartimento di Fisica, Universit$\bar{\mbox{a}}$ di Bologna,}}

\vspace*{10mm}
\centerline {\it Department of Physics, Syracuse University,}
\centerline {\it Syracuse, NY 13244-1130}
\vspace*{15mm}

\begin{abstract}

The exact quantization of two models, the massive vector meson model and the
Higgs model in the London limit, both describing massive photons, is presented.
Even though naive arguments (based on gauge-fixing) may indicate the
equivalence of these models, it is shown here that this is not true in general
when we consider these theories on manifolds with boundaries.  We show, in
particular, that they are equivalent only for a special choice of the boundary
conditions that we are allowed to impose on the fields.

\end{abstract}

\newpage
\baselineskip=24pt
\setcounter{page}{1}

\sxn{Introduction}\label{se:1}

\indent
It is known that gauge theories involve only massless gauge bosons unless
either the gauge symmetry is spontaneously broken \cite{Broke} or
topological terms \cite{CS,MCS} are added to
the action.  Such models are not merely of theoretical interest.  They have
important applications in particle physics and in condensed matter
physics.  In the former they arise, for example, in models that
incorporate the electroweak interactions \cite{Broke}.
Applications in condensed
matter physics arise in situations such as superconductivity
\cite{Super} where it is known that photons acquire a mass in the
superconducting regime.  Such models also arise in effective theories
describing the long wavelength physics \cite{Zee} of 2+1 dimensional systems.

It is usually believed that, under certain approximations, such
theories involving massive gauge bosons are equivalent to the massive
vector meson model.  By the massive vector meson model is here meant the
action whose gauge symmetry has been {\it explicitly}
broken by the addition of a quadratic mass term \cite{Massive}.
The arguments that are used \cite{Broke}
to show the equivalence of these models depend quite crucially on gauge-fixing
in some form or the other.  On the other hand, we know that gauge-fixing
arguments for manifolds with boundaries are always suspect because of
the following two reasons.

Firstly, the Gauss law for manifolds with boundaries has to be defined by
smearing it with appropriate test functions so as to ensure that they generate
canonical transformations \cite{Gauge} on the phase space.  Such a requirement
restricts the allowed gauge transformations \cite{Gauge} which, by
definition, are the canonical transformations generated by Gauss law.
Since gauge-fixing arguments do not usually pay attention to this feature,
they are not to be believed without further justifications.
The second reason is a subtle one and is related to the fact that
on a manifold with boundary, the Hamiltonian is self-adjoint and bounded
from below only if the fields and their momenta satisfy
suitable boundary conditions (BC's).  Most of the gauge-fixing arguments either
pick a special choice of BC's or do not talk about it at all.

Because of the above reasons, it is clear that the equivalence of the two
models is proven only
on manifolds without boundaries (like {\bf R$^{n}$}).  For manifolds with
boundaries,  a more careful analysis is
warranted to display the similarities/differences between the massive vector
meson model and a gauge theory with massive gauge bosons.

In Section 2, we look at the exact treatment (following Dirac) of the massive
vector meson model.
It is shown that the quantization here depends on a one-parameter family of
BC's.  We specialize to a particular BC for which we are able to carry out the
quantization completely.  We are not able to do this exact quantization for the
most general BC's.
In Section 3, we consider the Higgs model in the London limit (the
modulus of the Higgs field is frozen to its vacuum expectation value).
Here the exact treatment leads to a quadratic Hamiltonian
along with a Gauss law. In this case, quantization depends on a two-parameter
family of BC's and unlike the earlier case, we
are able to carry through the exact quantization for the most general BC's.
In Section 4, we compare the quantizations carried out in Sections 2 and 3.  It
turns out that there is a natural identification of the BC's of sections 2 and
3 provided one of the two parameters of section 3 is set to zero.  We show
that the Hamiltonian of the massive vector meson and Higgs models
are different in general.  In this Section, we
also briefly compare these two theories with one other
model describing a massive photon, namely the Maxwell-Chern-Simons
(MCS) \cite{CS,MCS} theory.

\sxn{The Massive Vector Meson Model} \label{se:2}
\indent
In this section we will consider the usual Maxwell action augmented by a
mass term for the vector potential $A_\m$ on a space time manifold $D
\times {\bf R}^1$, the two dimensional disc $D$ representing the spatial
manifold and ${\bf R}^1$ denoting time:
\be
S =  \int_{D \times {\bf R}^1}d^{3}x \left\{ - \frac{1}{4 e^2}\fmd \fmu
- \frac{1}{2} \ma ^2 \amd \amu \right\} \; .
\label{ml}
\ee
Here $F_{\m\n} = \pa_\m A_\n - \pa_\n A_\m$ is the electromagnetic
tensor whose components are the electric field $E_i = \pa_0 A_i -
\pa_i A_0$ and the magnetic field $B = \pa_1 A_2 - \pa_2 A_1 =
\frac{1}{2} \e_{ij} F_{ij}$
\footnote[1]{Throughout the paper we will
use the three-dimensional metric $\h$ with $\h_{00}=-1$ ,
$\h_{11}=\h_{22}=+1$ and the three dimensional Levi-Civita symbol
$\e_{\l\m\n}$ with $\e_{012}=+1$ .}.

It is well known \cite{Massive} that this model describes a
constrained system, with the second class constraints given by
\be
\P_0 \approx 0 \;\; \mbox{ and } \;\; \ma ^2 A_0 + \pa_i \Pi _i
\approx 0 \; , \label{con}
\ee
$\P_\m = (\P_0,\Pi _i)$ being the momenta conjugate to $A_\m$.  $\Pi _{i}$ here
is related to $E_{i}$ by
$\Pi _{i} = \frac{1}{e^{2}}E_{i}$.
Of course, the above constraints (\ref{con}) have to be smeared with
appropriate test functions so that they generate well-defined
canonical transformations \cite{Gauge}.  However, since these constraints are
second class, they can be imposed strongly.  Following Dirac's procedure
\cite{Dirac} for the
system described by (\ref{ml}), the Hamiltonian (up to irrelevant surface
terms that depend on the test functions used to smear the above constraints)
we end up with is
\be
H = \frac{1}{2} \int_{D} d^{2}x\left\{ e^2 \Pi _{i} ^{2} + \frac{1}{\ma^2}
(\pa_i \Pi _{i})^{2} + \frac{1}{e^2} B^2 + \ma^2 \aid^2
\right\} \; ,
\label{mh}
\ee
where the variables $A_i$ and $\Pi _i$ satisfy the usual canonical commutation
relations.
For the following discussion it is convenient to rewrite the
Hamiltonian (\ref{mh}) using the notation of differential forms
\footnote[2]{We write the fields $(A_1 ,A_2)$ , $(\Pi _1 ,\Pi _2)$ as the
one-forms $A=A_1 dx^1 + A_2 dx^2$ , $\Pi =\Pi _1 dx^1 + \Pi _2 dx^2$
respectively and the fields $B$ , $\pa_i \Pi _i$ as $*dA (=
\frac{1}{2}\epsilon ^{ij} F_{ij}$ , $-*d*\Pi$ respectively. In the latter
expression $*$ denotes the Hodge operation \cite{Hodge}.}.
After integrating by parts the
second and third terms in (\ref{mh}) and neglecting the surface
integrals, we get:
\be
H = -\frac{1}{2} \int_{D} \left\{ \Pi \, *(e^2 + \frac{1}{\ma^2}
d*d*) \Pi - A \, *(\ma^2 + \frac{1}{e^2} *d*d) A \right\} \; .
\label{fh}
\ee

We can rewrite this expression in a very compact form if on the vector
space of forms $\a^{(p)}$ of degree $p$ we introduce the scalar product
$<\a^{(p)},\b^{(p)}> \, := (-1)^p \int \bar{\a^{(p)}} \, *\b^{(p)}$,
where the bar denotes complex conjugation. Now (\ref{fh}) becomes
\be
H = \frac{1}{2} <\Pi ,(e^2 + \frac{1}{\ma^2}d*d*) \Pi > +
\frac{1}{2} <A,(\ma^2 + \frac{1}{e^2} *d*d) A> \; .
\label{sh}
\ee

In order to quantize this Hamiltonian, we need to expand the fields
$A$ and $\Pi$ in a complete basis of the Hilbert space of one-forms.
Since $H$ is constructed from the differential operators $d*d*$ and
$*d*d$, we would like to expand the fields in a basis of
eigenfunctions of such operators \cite{MCS}.  In other words, we need to find a
domain of self-adjointness\footnote[3]{Let us recall that the property
defining the domain
$\cd \subset \ch$ of a self-adjoint operator $T$ on a Hilbert space
$\ch$ is the following \cite{Domain}: $<\chi,T\h>-<T\chi,\h> = 0$ , $\forall
\h \in \cd \Leftrightarrow \chi \in \cd$.} for these operators.  Let us
first find a domain of self-adjointness \cite{Domain} for the operator
$*d*d$.  From the relation:
\be
0 = <\a,*d*d \b> - <*d*d \a, \b> = \int_{\pa D}
\left\{ \bar{*d\a}
\, \b - \bar{\a} \, *d \b \right\} \; ,\label{sa}
\ee
where $\a,\b$ are any two one-forms, it is easy to see that the
operator $*d*d$ is self-adjoint on the domain
\be
\cd_\l = \left\{ \a \; : \; *d\a \, |_{r=R} = - \l \a_{\q} \, |_{r=R}
\right\} \;\;\;\; \l \in {\bf R}^1 \; .
\label{do}
\ee
To go from (\ref{sa}) to (\ref{do}) we have required that the fields
satisfy local rotationally invariant BC's.  [By local BC's, we mean
BC's which mix fields and their derivatives only at the same point.]
In (\ref{do}), $r$ and $\q$ are
polar coordinates on the disc $D$ with $r=R$ giving its boundary and
$A_\q = A_i \frac{\pa x^i}{\pa \q} (r,\q)$.

$\l$ here can be any real number. But the
Hamiltonian (\ref{sh}) is bounded from below only if $\l \geq 0$.
This can be seen by noticing that
\be
<\a ,*d*d \a > = <d \a ,d \a > - \int_{\pa D} \bar{\a} (*d \a)
= <d \a ,d \a > + \l \int_{\pa D} |\a_\q |^2 R d\q \; . \label{po}
\ee
Since $<d \a ,d \a > \geq 0$, $<\a ,*d*d \a >$ is
nonnegative iff $\l \geq 0$. Therefore, from now on, we will
only consider the domains $\cd_\l $ with $\l \geq 0$.
Let us now turn to the problem of solving the eigenvalue equation
\be
*d*d A = \o^2 A
\label{ep}
\ee
for the one-form $A$ satisfying the BC's
\be
*d A \, |_{r=R} = - \l A_\q \, |_{r=R} \; \; , \; \; \l \geq 0
\; . \label{bc}
\ee
This problem has already been examined and solved in \cite{MCS}. Here,
we will not repeat the calculations and will only list the
eigenmodes of $*d*d$, together with the corresponding eigenvalues
$\o ^2$.

The solutions for $\o ^2 \neq 0$ are of the form:
\be
\left.
\begin{array}{l}
\pnm^{(1)} = \cn_{nm}^{(1)}\; *d[e^{in\q} J_n (\wn^{(1)} r)] ~~~~\\
\Psi_{-nm}^{(1)} = \bar{\pnm^{(1)}}
\end{array}
\right\} n \geq 0 \, , \, m > 0 \; ,
\label{m1}
\ee
where $J_n (x)$ is the real Bessel function of order $n$\footnote{Here and in
the following, we adopt the convention that
the normalization contants, such as $\cn_{nm}^{(1)}$ in (\ref{m1}),
are fixed by
the conditions $<\Psi_{nm},\Psi_{nm}>=1$ and $\cn_{nm}^{(1)} > 0$.} and the
eigenvalues $\o = \o_{nm}^{(1)}$ are fixed by the BC's (\ref{bc}) which now
read:
\be
\wn^{(1)} J_n (\wn^{(1)} R) = \l \left[ \frac{d}{d(\wn^{(1)} r)}
J_n(\wn^{(1)} r) \right]
_{r=R} \; .
\label{bcm1}
\ee

There is also a set of zero modes,
solutions of (\ref{ep}) with $\o =0$, given by:
\be
\left.
\begin{array}{l}
\ponm =  \cn_{nm}^{(0)}\; d[e^{in\q} J_n (\wn^{(0)} r)] ~~~~\\
\Psi_{-nm}^{(0)} = \bar{\ponm}
\end{array}
\right\} n \geq 0 \, , \, m > 0 \; ,
\label{m0}
\ee
the Bessel functions now satisfying the condition $J_n (\wn^{(0)}R) = 0$.

The functions (\ref{m1}) and (\ref{m0}) form a complete set of
solutions for (\ref{ep}),(\ref{bc}) if $\l >0$. On the contrary, if $\l
=0$ there is another set of zero modes, given by the so-called
harmonic forms
\begin{eqnarray}
\hn = \cn_n^{(h)}\; d z^n \;\;\; ,\;\;\; \bar{\hn} = \cn_n^{(h)} \; d \bar{z}^n
&& n >0 \label{hm}
\end{eqnarray}
where $z= x_1 + i x_2 = r e^{i\q}$ is the complex coordinate on
$D$.

It is important to notice that the harmonic functions are eigenfunctions
of the \\ operator *, $*h_n = i h_n$,
and that, in addition, if $\l =0$ there exists a
relationship between the nonzero modes (\ref{m1}) and the zero modes
(\ref{m0}) $\pnm^{(1)} = *\ponm \; \mbox{ with }\; \o_{nm}^{(1)} = \o_{nm}
^{(0)}$.
These two relations imply that, for $\l =0$, the complete set of
eigenfunctions $(\Psi_{nm}^{(1)},\Psi_{nm}^{0},h_n )$ of the operator
$*d*d$ is also a complete set of eigenfunctions of the operator
$d*d*$.   Indeed, to diagonalise $H$, we can expand the fields $A$ and $\Pi$
in (\ref{sh}) as \bea
A &=& \qn^{(1)} \pnm^{(1)} + \qon \pnm^{(0)} + q^{(h)}_n \hn + c.c.
\; ,\nonumber \\
\Pi &=& \pn^{(1)} \pnm^{(1)} + \pon \pnm^{(0)} + p^{(h)}_n
\hn + c.c. \; ,\label{exp}
\eea
where (as in the following) repeated indices are summed over.  The Hamiltonian
is then
\bea
H &= & \frac{1}{2} \left\{ \left[ \left( e^2 +
\frac{\wn^{2}}{\ma^2} \right) \pn^{(0)\dagger} \pn^{(0)} + \ma^2
\qn^{(0)\dagger} \qn^{(0)} \right] + \left[ e^2 \pn^{(1)\dagger} \pn^{(1)} +
\left( \frac{\wn^{2}}{e^2} + \ma^2 \right) \qn^{(1)\dagger} \qn^{(1)}\right] +
\right. \nonumber \\
& + & \left.
\left[ e^2 p^{(h)\dagger}_n p^{(h)}_n + \ma^2 q^{(h)\dagger}_n q^{(h)}_n
\right] \right\} \; ,
 \label{dh}
\eea
where the only non zero commutation relations satisfied by the operators
$q^{(j)}$'s and $p^{(j)}$'s ($j=0,1,h$) are $
\left[ \qn^{(1)},p_{n'm'}^{(1)\dagger} \right] = i \d_{nn'} \d_{mm} = \left[
\qon,p_{n'm'\dagger}^{(0)} \right] \;\; , \;\;
\left[ q^{(h)}_n ,p^{(h)\dagger}_{n'} \right] = i \d_{nn'}$.

Here $\omega_{nm}=\omega _{nm}^{(0)}=\omega _{nm}^{(1)}$ while
the commutation relations follow from those of the variables $A_{i},\Pi _{i}$.
  Let us define the annihilation-creation operators
\be
 \left.
\begin{array}{l}
a_{nm}^{(j)} = \frac{1}{\sqrt{2}} \left[ \frac{1}{\sqrt{\O_{nm}^{(j)}}}
            \pn^{(j)} - i \sqrt{\O_{nm}^{(j)}} \qn^{(j)}  \right] ~~~~\\
a_{nm}^{(j)\dagger} = \frac{1}{\sqrt{2}} \left[
\frac{1}{\sqrt{\O_{nm}^{(j)}}}
            \pn^{(j)\dagger} + i \sqrt{\O_{nm}^{(j)}}
            \qn^{(j)\dagger}  \right] ~~~~
\end{array} \right\} j=0,1   \label{ca0}
\ee
\be
\begin{array}{l}
 a_{n}^{(h)} = \frac{1}{\sqrt{2}} \left[ \frac{1}{\sqrt{\O_{n}^{(h)}}}
            p^{(h)}_n - i \sqrt{\O_{n}^{(h)}} q^{(h)}_n  \right] \\
a_{n}^{(h)\dagger} = \frac{1}{\sqrt{2}} \left[
\frac{1}{\sqrt{\O_{n}^{(h)}}}
            p^{(h)\dagger}_n + i \sqrt{\O_{n}^{(h)}}
            q^{(j)\dagger}_n  \right] \end{array}
 \label{ca} \; ,
\ee
with commutators $
\left[ a_{nm}^{(j)} , a_{n'm'}^{(j)\dagger} \right] =
i \d_{nn'} \d_{mm'} \;\; (j=0,1)\;\; , \;\;
\left[ a_{n}^{(h)} , a_{n'}^{(h)\dagger} \right] =
i \d_{nn'}$,
where
\be
\O_{nm}^{(1)}=\O_{nm}^{(0)}=\sqrt{\wn^{(0)2} + e^2 \ma^2} \;\; , \;\;
\O_{n}^{(h)} = e \ma \; .
\label{om}
\ee

Then (\ref{dh}) becomes
\be
H = \O_{nm}^{(1)} \a_{nm}^{(1)\dagger} \a_{nm}^{(1)} +
\O_{nm}^{(0)}
\a_{nm}^{(0)\dagger} \a_{nm}^{(0)} + \O_{n}^{(h)} a_n^{\dagger} a_n \; .
\label{hd}
\ee
The spectrum of $H$ can be read off from (\ref{hd}).

We remark here that the lowest energy modes are the ones
corresponding to the harmonic functions and are all degenerate,
having an energy $\O_{n}^{(h)}$ that depends only on the mass $\ma$ of
the vector potential, for all $n$.

We conclude this section by looking briefly at the $\l >0$ case.  Now it is
no longer true that the operator $d*d*$ is diagonal in the basis
$(\Psi_{nm},\Psi_{nm}^{(0)})$ of eigenfunctions of $*d*d$.  If we continue to
use a field expansion similar to (\ref{exp}), with the harmonic modes now
missing, we can write (\ref{sh}) in the non-diagonal form
\bea
H & = & \frac{1}{2} \left\{ \left[ \left( e^2  \d_{mm'} +
\frac{\omega _{nm'}^{(0)2}f_{mm'}}{\ma^2} \right) p_{nm}^{(0)\dagger}p_{nm'}
^{(0)}
+ \ma^2 \qn^{(0)\dagger} \qn^{(0)} \right] \right. + \nonumber\\
& & \left. \left[ e^2 \pn^{(1)\dagger}
\pn^{(1)}+\left( \frac{\wn^{(1)2}}{e^2} + \ma^2 \right) \qn^{(1)\dagger}
\qn^{(1)} \right] \right\}
\label{ndh}
\eea
where the overlap coefficients $f_{mm'}:= <\Psi_{nm}^{(1)} , *\Psi_{nm'}^{(1)}>
$ are different from zero
for every $m,m'$. Thus, each of the $\pn^{(0)}$ mode of the momentum
field is coupled to an infinite number of other such modes. We do not
know how to diagonalize (\ref{ndh}).

\sxn{The Higgs Model} \label{se:3}

As before, we will work on the space-time manifold $D
\times {\bf R}^1$. Consider then a $U(1)$ Higgs model with the modulus
$\r$ of the Higgs field $\f = \r e^{iq\j}$ frozen to its vacuum value. In
this limit (the London limit), in addition to the vector potential
$A_\m$, the only other degree of freedom is the real phase $\j$.  The action in
these variables is then:
\be
S = \int_{D \times {\bf R}^1} d^{3}x\left\{ -\frac{1}{4 e^2} \fmd F^{\mu\nu}-
\frac{m^2_{H}}{2} (\pa_\m \j - \amd) (\pa^\m \j - \amu) \right\}
\label{hl}
\ee
where $m{H}e = q \r$ is the mass of the vector meson $A_\m$.

The Hamiltonian corresponding to (\ref{hl}) is
\be
H = \int_{D} d^{2}x\left\{ \frac{e^{2}}{2} \Pi _{i} ^2
+ \frac{1}{2e^{2}} B^2 + \frac{1}{2m^2_{H}} \P^2 +
\frac{m^2_{H}}{2} (\pa_i \j - \aid)^2 \right\} \label{hh}
\ee
where $\Pi _{i}=\frac{E_{i}}{e^{2}}$ is the momentum conjugate to $A_{i}$ and
$\Pi$ the momentum conjugate to $\psi$.  This Hamiltonian has to supplemented
by the Gauss law $\P - \pa_i \Pi _{i} \approx 0$.
The non-zero Poisson Brackets are
\be
\left\{ \j(x), \P(y) \right\} = \d^2 (x-y) \;\; , \;\;
\left\{ A_i (x) , \Pi _j (y) \right\} =  \d_{ij} \d^2 (x-y) \; .
\label{hc}
\ee
As explained in \cite{Gauge} and after equation (\ref{con}), the correct way
of reading the Gauss law is by smearing it with a test function $\L^{(0)}$:
\be
\cg (\L^{(0)}) = \int_{D} d^{2}x\L^{(0)} \left( \P - \pa_i \Pi _{i} \right) =
0 \; , \label{sg}
\ee
where $\L^{(0)}$ is zero on the boundary of the disc,
$\L^{(0)}|_{\pa D} = 0$.

Let us rewrite both (\ref{hh}) and (\ref{sg}) using the form
notation. To do so, let us introduce the space of vectors $(\j ,A)$
where $\j$ is a zero-form and $A$ is a one-form, with the scalar product
$<(\j ,A),(\j' ,A')> = <\j,\j'>_0 + <A,A'>_1$
where $< \cdot , \cdot >_0$ , $< \cdot , \cdot >_1$ are
the scalar products respectively on zero and one forms, as
previously defined.
In addition, to simplify the notation, we set $f := m_{H} \j$,
$P := \P /m_{H}$, ${\cal A}_{i}:= A_{i}/e$ and ${\cal E}_{i}:=
e\Pi _{i}$.

After integrating (\ref{hh}) by parts and neglecting surface terms,
the Hamiltonian becomes
\bea
H &=& \frac{1}{2}( <(f,{\cal A}), \hat{H}_0 (f,{\cal A})>+<(P,{\cal
P}),(P,{\cal E})>) \label{ch} \\
\hat{H}_0 &=& \left[ \begin{array}{cc}
*d*d & -m_{H}e *d*  \\ -m_{H}e d & *d*d +m ^2_{H} e ^2 \end{array} \right]
\nonumber
\eea
while the Gauss law (\ref{sg}) reads
\be
\cg(\L^{(0)}) = <\L^{(0)},P +\frac{1}{m_{H}e} *d* {\cal E}> = 0 \; .
\label{cg}
\ee

Our analysis will now proceed as in the massive vector meson case. We
have first to look for suitable BC's on the fields $(f,{\cal A})$ that make
the Hamiltonian $H$ diagonalizable. Let $f,g$ be zero-forms and ${\cal A},{\cal
B}$ be one-forms.  Then, from
\bea
&&<(g,{\cal B}),\hat{H}_0 (f,{\cal A})> - <\hat{H}_0 (g,{\cal B}),(f,{\cal A})>
= \nonumber \\
&&
= \int_{\pa D} \left\{ \bar{*d {\cal B}} \,{\cal A} - \bar{{\cal B}} \, *d
{\cal A}
\right\} - \int_{\pa D} \left\{ \bar{g} \, *(df - m_{H}e{\cal A})
- \bar{*(dg-m_{H}e{\cal B})} \, f \right\},
\label{st}
\eea
we see that $\hat{H}_{0}$ is self-adjoint on the
domain
\be
\cd _{\l \m} = \left\{ (f,{\cal A}) \; : \; *d{\cal A} \, |_{r=R}
= - \l {\cal A}_\q \, |_{r=R} \mbox{ and }
f \, |_{r=R} = -\m [*(df-m_{H}e{\cal A})]_\q \, |_{r=R} \right\} \;\;\;
\l ,\m \in {\bf R}^1 . \label{hdo}
\ee
We note here that $\lambda$ has the same role as the $\lambda$ that appeared in
section 2, while $\mu$ is a new parameter that did not exist in the previous
case.  We have as before imposed locality and rotational invariance to
obtain (\ref{hdo}).

On this domain, the Hamiltonian can be rewritten as
\bea
H  &=& \frac{1}{2} \left\{ <{\cal E},{\cal E}>_1 + <P,P>_0 + <d{\cal A},d{\cal
A}>_1 + <df-m_{H}e{\cal A},df-m_{H}e{\cal A}> \right\} + \nonumber \\
&+& \frac{1}{2} \int_{\pa D} R
d\q\left\{  \l |{\cal A}_\q |^2 + \frac{1}{\m} f^{2} \right\}_{r=R} \; ,
\label{hpd}
\eea
so that it is nonnegative iff $\l ,\m \geq 0$. Thus from now
on we will consider only domains $\cd_{\l \m}$ with $\l \geq 0 \; ,
\mu \geq 0$.

Our task then is to solve the eigenvalue problem
\be
\hat{H}_0 \left( \begin{array}{c} f \\ {\cal A} \end{array} \right)
= \o^2  \left( \begin{array}{c} f \\ {\cal A} \end{array} \right)
\label{eph}
\ee
with the BC's
\bea
 *d{\cal A} \, |_{r=R} &=& - \l {\cal A}_\q \, |_{r=R} \label{bch1} \\
f \, |_{r=R} &=& -\m [*(df-m_{H}e{\cal A})]_\q  \, |_{r=R} \; . \label{bch2}
\eea
The eigenmodes of (\ref{eph}) subject to the above BC's can be found by an
analysis very similar to that used to solve (\ref{ep}).

If $\o^2 \neq 0$, this system of equations can be decoupled into one
differential equation for ${\cal A}$ and one equation defining $f$ as a
function of ${\cal A}$:
\bea
(d*d* + *d*d) {\cal A} &=& (\o^2 -m^2_{H} e^2 ) {\cal A} \; ,\label{ea} \\
f &=& -\frac{1}{m_{H}e} *d*{\cal A} \; . \label{ef}
\eea

Let us first look at the modes of ${\cal A}$ obtained from equation
(\ref{ea}) when $\o^2 = m^2_{H} e^{2}$. In this case the harmonic one forms
(\ref{hm}) satisfy (\ref{ea}) and the BC (\ref{bch1}) if $\l = 0$.
In addition, from (\ref{ef}) and (\ref{hm}) it follows that $f \equiv 0$,
so that (\ref{bch2}) is satisfied only for $\m =0$. This means that the
Hamiltonian (\ref{ch}) admits harmonic modes for $\l =0$ iff $\m$ is also zero.
Since we are interested in comparing the Higgs
model in the broken phase with the massive vector meson model and
since the latter does admit harmonic modes for $\l =0$, from now on
we will set $\m \equiv 0$.

Thus, if $\l =0$, we have a set of solutions $(f,{\cal A})$ of (\ref{eph})
corresponding to the eigenvalue $\o ^2 = m^2_{H} e^{2}$ given by
\begin{eqnarray}
\Hn = \cm_n^{(h)}\; \left( 0,d z^n \right) \;\;\; ,\;\;\; \bar{\Hn} =
\cm_n^{(h)}\; \left( 0,d \bar{z}^n \right)&&
n >0 \; . \label{hhm}
\end{eqnarray}

If $\o ^2 > m^2_{H} e^{2}$, (\ref{ea}) and (\ref{ef}) admit the following two
sets of solutions for $\l \geq 0$:
\be
\left.
\begin{array}{l}
\fnm^{(\a )} = \cm_{nm}^{(\a )} \left( -\frac{1}{m_{H}e} e^{in\q} J_n(\a_{nm}
r) ,\frac{1}{\a_{nm}^2} d[e^{in\q} J_n (\a_{nm} r)] \right) ~~~~\\
\F_{-nm}^{(\a )} = \bar{\fnm^{(\a )}}  \end{array}
\right\} n \geq 0 \, , \, m > 0 \; ,
\label{saa}
\ee
\be
\left.
\begin{array}{l}
\fnm^{(\b )} = \cm_{nm}^{(\b )} \left( 0
,\frac{1}{\b_{nm}^2} *d[e^{in\q} J_n (\b_{nm} r)] \right) ~~~~\\
\F_{-nm}^{(\b )} = \bar{\fnm^{(\b )}}  \end{array}
\right\} n \geq 0 \, , \, m > 0 \; .
\label{sab}
\ee
corresponding to the eigenvalues $\o^{(\a )2}_{nm} = \a_{nm}^2 +
m^2_{H}e^{2}$ and $\o^{(\b)2}_{nm}=\b_{nm}^2 +m^2 _{H}e^{2}$ respectively,
where
$\a_{nm}$ and $\b_{nm}$ are determined by the BC's (\ref{bch1}) and
(\ref{bch2}) with $\m =0$:
\bea
J_n (\a_{nm}R) &=& 0 \; ,\label{bc11} \\
\b_{nm} J_n (\b_{nm}R) &=& \l \left[ \frac{d}{d(\b_{nm} r)}
J_n(\b_{nm} r) \right] _{r=R} \; .\label{bc12}
\eea
(Notice that the $\a _{nm}$ and $\b _{nm}$ above are respectively
identical to the $\omega _{nm}^{(0)}$ and $\omega _{nm}^{(1)}$ of Section 2.)

Finally we obtain a set of solutions to (\ref{eph}) when $\o^2 =0$, which
are given by:
\be
\left.
\begin{array}{l}
\F_{nm}^{(0)} = \cm_{nm}^{(0)} \left( e^{in\q} J_n (\g_{nm}r),
\frac{1}{m_{H}e} d[e^{in\q} J_n (\g_{nm}r)] \right) ~~~~\\
\F_{-nm}^{(0)}=\bar{\F_{nm}^{(0)}} \end{array}
\right\} n \geq 0 \, , \, m > 0 \; ,
\label{sao}
\ee
with the $\g_{nm}$'s fixed by the condition $J_n (\g_{nm} R) = 0$ (so that
$\g_{nm}=\a_{nm}$).

In conclusion, for $\l =0$, (\ref{hhm},\ref{saa},\ref{sab},\ref{sao}) form
a   complete set of eigenfunctions that allow us to expand the fields
$(f,{\cal A})$ and $(P,{\cal E})$ as
\bea
(f,{\cal A}) &=& \qn^{(\a )} \fnm^{(\a )} + \qn^{(\b )} \fnm^{(\b )} +\qon
\fonm +
q^{(h)}_n \Hn + c.c. \nonumber \\
(P,{\cal E}) &=& \pn^{(\a )} \fnm^{(\a )} + \pn^{(\b )} \fnm^{(\b )} +\pon
\fonm +
p^{(h)}_n \Hn + c.c.
\label{hexp}
\eea
where, by virtue of (\ref{hc}), the only nonzero commutation relations
are $\left[ q_{nm}^{(j)} , p_{n'm'}^{(j)} \right] = i \d_{nn'} \d_{mm'}
\; \; , \; \;
\left[ q_{n}^{(h)} , p_{n'}^{(h)} \right] = i \d_{nn'}$,
where $j=\alpha ,\beta$ or $0$.

If $\l >0$, the field expansions look very similar to (\ref{hexp}),
except for the fact that in this case the harmonic modes $H_n$ are
absent.
We now will turn our attention to the Gauss law (\ref{cg}). Since in
(\ref{cg}) the test function $\L^{(0)}$ vanishes on the boundary $\pa
D$, it can be chosen in particular to be $e^{in\q} J_n (\g_{nm} r)$
with $J_n (\g_{nm} R)=0$. It is then immediate
to verify that Gauss law simply implies $\pn^{(0)} \approx 0$.

We can also introduce creation-annihilation operators as in
(\ref{ca0},\ref{ca}),
where now $j=\a ,\b$, and
\be
\O_{nm}^{(\a )}= \sqrt{\a^2_{nm} + m^2_{H} e^{2}} \;\;,\;\;
\O_{nm}^{(\b )}= \sqrt{\b_{nm}^2 + m^2_{H} e^{2}}\;\; ,\;\;
\O_{n}^{(h)} = m_{H}e \; . \label{omm}
\ee

Therefore, the Hamiltonian (\ref{ch}) acting on the physical states becomes:
\be
H = \O_{nm}^{(\a )} a_{nm}^{(\a )\dagger} a_{nm}^{(\a )} +
\O_{nm}^{(\b )} a_{nm}^{(\b )\dagger} a_{nm}^{(\b )} +
\O_{n}^{(h)} a_n^{(h)\dagger} a_n^{(h)} \; ,
\label{hdd} \ee

Notice that the Hamiltonian (\ref{hdd}) has been explicitly derived
from (\ref{ch}) for $\l =0$. In this case, $\Omega _{nm}^{(\alpha )}=\Omega
_{nm}^{(\beta )}=\Omega _{nm}$ just as in the massive vector meson case for
$\lambda =0$.  For $\l >0$, (\ref{ch}) assumes a
form which is almost identical to (\ref{hdd}), the only difference being that
the harmonic modes are no longer solutions of (\ref{ea}) and hence do not
appear in the Hamiltonian.

\sxn{Conclusions}\label{se:4}

Let us now compare the vector meson model with the Higgs model
in the London limit.
{}From (\ref{exp}) and (\ref{hexp}) it is clear that there is a
one-to-one correspondence between the modes of the fields for these
two theories (if we take into account the Gauss law $p_{nm}^{(0)}\approx 0$
which
kills one set of modes for the Higgs theory). But what about the Hamiltonians?

It is known \cite{Broke} that on an infinite
plane these two models are
indeed equivalent, both describing a massive electromagnetic
potential $A_\m$. We see this also in our approach, by noting
that the only BC's that are suitable to a plane geometry
require that all the fields vanish at infinity and hence force both
$\l$ and $\m$ to be zero. In the latter case, the Hamiltonian for the
massive vector meson model (\ref{hd}) and the one for the Higgs model
(\ref{hdd}) are exactly the same, once we identify $m_A$ with $m_H$. Therefore,
both on an infinite plane and on a disc with
BC's $\l = \m =0$, these two models coincide.

This is not the case if we confine the theory on a disc and impose
BC's with $\l >0$ (but still $\m =0$). The two Hamiltonians
are then different: while for the Higgs model, (\ref{hdd}) is
diagonal, in the massive vector meson model it has the form
(\ref{ndh}), in which every mode of the electric field is coupled to
infinitely many others.
Thus the massive vector meson model and the Higgs model are equivalent on
a disc only if we choose boundary conditions for the fields
characterized by the value zero for the parameter(s) that appear in
(\ref{bc}) and (\ref{bch1},\ref{bch2}).

We would like to end this paper by briefly comparing the models under
consideration to yet another model describing a massive vector
meson, namely the Maxwell-Chern-Simons (MCS) theory. The action for this model
reads:
\be
S = \int_{D\times {\bf R}^1 }d^{3}x
\left\{ -\frac{1}{4 e^2} F_{\m\n} F^{\m\n} +
\frac{k}{4\pi} \epsilon^{\mu\nu\rho} A_{\mu}
\partial_{\nu} A_{\rho} \right\} \; .
\label{mcs}
\ee

This Lagrangian has been studied in detail in \cite{MCS}, where it has been
shown that
the fields have to satisfy BC's characterized by a nonnegative
parameter $\l$, exactly like in (\ref{bc}) or (\ref{bch1}). As in
section 2, one can show that the Hamiltonian of this system is
diagonalized by the modes of the operator $*d*d$ only if $\l =0$ and
that, as soon as $\l$ deviates from this value, the Hamiltonian couples
an infinite number of such modes. In fact, the Hamiltonian for the
MCS theory derived in \cite{MCS} and the one for the massive vector
meson model, (\ref{ndh}), do coincide if we make the identifiction
$m_A = \frac{ek}{2\p}$. In particular, both these
Hamiltonians concide with the Higgs Hamiltonian (\ref{hdd}) when $\l$
is chosen to be zero.  In this latter case, the MCS theory admits an additional
set of observables (``edge'' observables) which commute with the Hamiltonian
and are completely localized at the boundary of the spatial manifold $D$.

{\bf Acknowledgements}
We thank A.P. Balachandran for many useful suggestions that he gave us
throughout this work.  We also thank P. Teotonio and A. Momen
for stimulating discussions.  This work was supported by the DOE, U.S.A.
under contract number DE-FG02-85ER40231.

\end{document}